%% file: main.tex
\documentclass[letterpaper, 10 pt, conference]{IEEEtran}

\IEEEoverridecommandlockouts

\usepackage[export]{adjustbox}
\usepackage{graphicx} 
\usepackage{epsfig}
\usepackage{pgfplots} 
\usepackage{booktabs}
\pgfplotsset{plot coordinates/math parser=false} 
\newlength\figureheight 
\newlength\figurewidth 

\usepackage{amsmath,amsfonts}
\usepackage{algorithmic}
\usepackage{algorithm}
\usepackage{array}
\usepackage{tikz}\usetikzlibrary{arrows}
\usepackage{subfig}
\usepackage{textcomp}
\usepackage{stfloats}
\usepackage{url}
\usepackage{verbatim}
\usepackage{amsmath,amsfonts}
\usepackage{amssymb}
\usepackage{gensymb}    
\usepackage{siunitx}
\usepackage{tikz}\usetikzlibrary{arrows}
\usepackage{enumitem}
\usepackage{theorem}

\usepackage{cite}

\begin{document}
\title{Learning-based Approximate Model Predictive Control for an Impact Wrench Tool}

\author{
    \IEEEauthorblockN{
        Mark Benazet\textsuperscript{1}, 
        Francesco Ricca\textsuperscript{1}, 
        Dario Bralla\textsuperscript{2}, 
        Melanie N. Zeilinger\textsuperscript{1},
         Andrea Carron\textsuperscript{1}
    }
    \thanks{\textsuperscript{1}Institute of Dynamic Systems and Control, ETH Zurich, Switzerland}
    \thanks{\textsuperscript{2}Hilti Corporation, Schaan, Liechtenstein}
}

\markboth{Journal of \LaTeX\ Class Files,~Vol.~14, No.~8, March~2024}%
{Shell \MakeLowercase{\textit{et al.}}: A Sample Article Using IEEEtran.cls for IEEE Journals}

\maketitle

\input{Sections/abstract}   
\input{Sections/keywords}
\input{Sections/Introduction}

\input{Sections/System_overview}
\input{Sections/Methodology}
\input{Sections/Numerical_Results}
\input{Sections/Conclusion}

\bibliographystyle{IEEEtran}
\bibliography{bibliography}

\end{document}

%% file: Sections/abstract.tex
\begin{abstract}
Learning-based model predictive control has emerged as a powerful approach for handling complex dynamics in mechatronic systems, enabling data-driven performance improvements while respecting safety constraints. However, when computational resources are severely limited, as in battery-powered tools with embedded processors, existing approaches struggle to meet real-time requirements. In this paper, we address the problem of real-time torque control for impact wrenches, where high-frequency control updates are necessary to accurately track the fast transients occurring during periodic impact events, while maintaining high-performance safety-critical control that mitigates harmful vibrations and component wear. The key novelty of the approach is that we combine data-driven model augmentation through Gaussian process regression with neural network approximation of the resulting control policy. This insight allows us to deploy predictive control on resource-constrained embedded platforms while maintaining both constraint satisfaction and microsecond-level inference times. The proposed framework is evaluated through numerical simulations and hardware experiments on a custom impact wrench testbed. The results show that our approach successfully achieves real-time control suitable for high-frequency operation while maintaining constraint satisfaction and improving tracking accuracy compared to baseline PID control.
\end{abstract}

%% file: Sections/keywords.tex

%% file: Sections/Introduction.tex
\section{Introduction}
\label{sec:1_Introduction}

\begin{figure}[h]
    \centering
    \includegraphics[width=0.5\linewidth, trim={0cm 0cm 0cm 0cm}, clip]{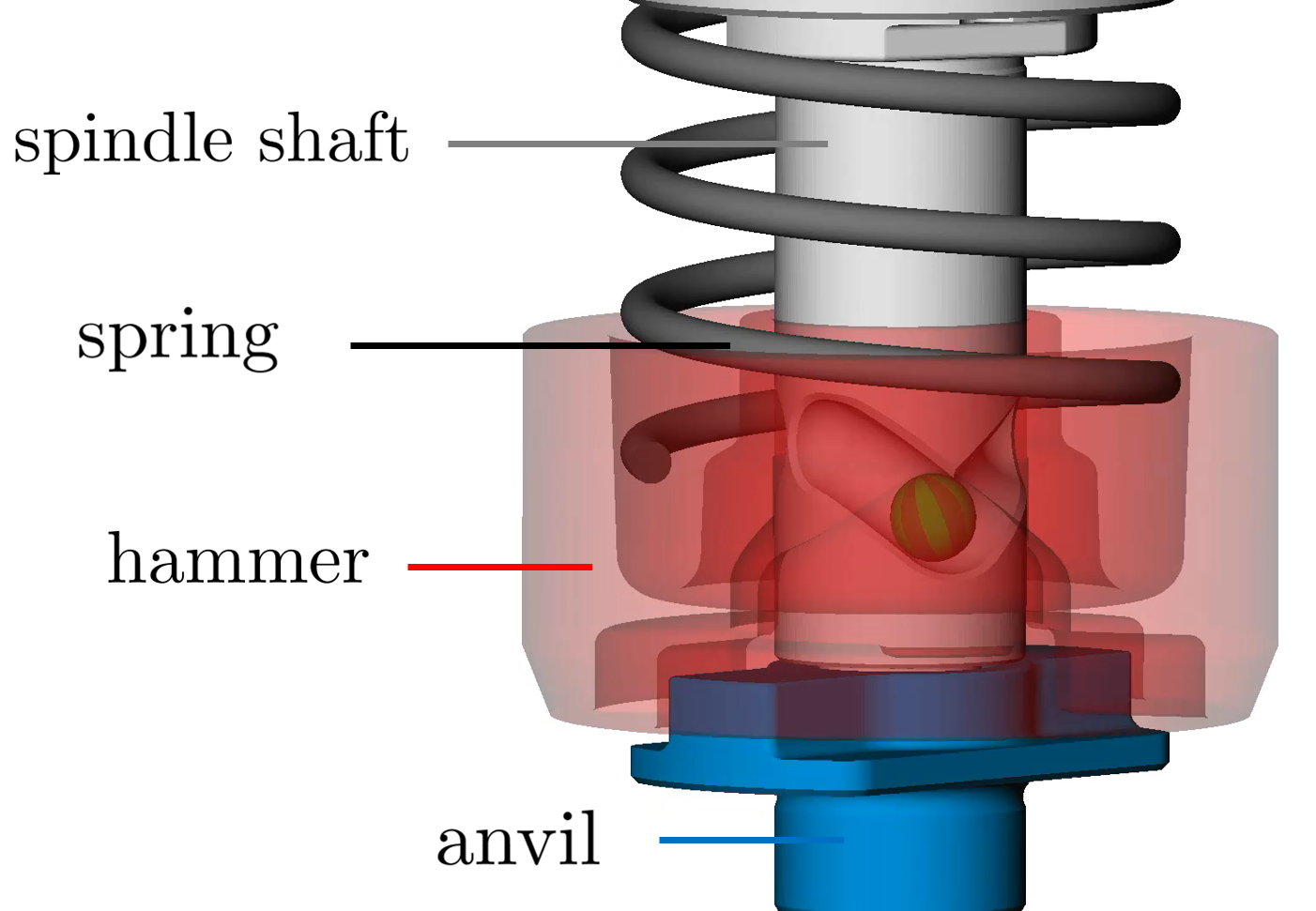}
    \caption{The internal mechanism of the impact wrench. The motor torque drives the spindle rotation, which transmits force to the hammer through the ball-and-cam mechanism. The spring stores and releases energy during the impact cycle.}
    \label{fig:internals}
\end{figure}

Impact wrenches are widely used power tools in construction, automotive, and manufacturing industries, designed to deliver high-torque outputs through rapid, periodic impacts. Figure \ref{fig:internals} shows the primary components of the impact wrench. The tool comprises a hammer, i.e., a rotating mass driven by a spindle shaft via a spring-ball mechanism, and an anvil on which a socket is mounted to drive fasteners. The working principle is to store rotational kinetic energy from the motor and release it in short bursts to the anvil. While this efficient design enables fast operation, it introduces significant control challenges. Operators are exposed to vibrations that pose potential safety hazards. Numerous studies have investigated the physical consequences of prolonged vibration exposure \cite{Hand-armVibrationEffectsonPerformanceTactileAcuityandTemperatureofHand, peripheralnerveendings} and developed monitoring techniques to mitigate health risks \cite{EvaluationofImpactWrenchVibrationEmissionsandTestMethods, InvestigCharacVibr}. Besides safety issues, vibrations can also cause damage to the components of the impact wrench. These issues can be mitigated by maximizing the hammer-anvil contact area at impact, which helps prevent high contact pressures and material fatigue while avoiding axial impacts that dissipate kinetic energy and reduce output torque.

Model predictive control (MPC) is a natural candidate for this application due to its ability to handle constraints and optimize performance over a prediction horizon. However, the computational burden of solving the nonlinear optimization problem at each control step is prohibitive for the embedded processor in the power tool, which must execute control updates at $1$ kHz. This high-frequency operation, combined with limited computing resources, renders standard MPC implementations infeasible. Additional challenges arise from parameter uncertainties due to manufacturing tolerances, as the MPC controller requires accurate parameter knowledge.

We address these challenges in this paper through a learning-augmented control architecture. We present a model predictive control (MPC) scheme augmented with Gaussian process (GP) regression to learn residual dynamics from experimental data. To meet real-time evaluation requirements, we approximate the resulting MPC control law with a neural network (NN). State estimation from partial measurements is performed using an Extended Kalman Filter (EKF).

\subsubsection*{Contributions}
The contributions of this paper are threefold: Firstly, we develop an MPC control scheme that accurately generates torque commands for an impact wrench system while explicitly handling safety and performance constraints. To this end, we augment the nominal dynamics model with Gaussian process-based residual learning from hardware data, which allows the controller to capture unmodeled effects directly from experimental observations. This data-driven augmentation is integrated within the MPC, which enables improved trajectory tracking and constraint satisfaction under real operating conditions. Secondly, we propose a neural network approximation of the GP-augmented MPC controller that achieves the required evaluation times on embedded hardware. In particular, we employ supervised learning on a diverse dataset of optimal MPC solutions to train a compact feed-forward architecture. Owing to this extensive training across varying operating conditions, the resulting neural network controller learns to replicate the MPC while achieving microsecond-level inference on resource-constrained embedded platforms. This enables high-frequency control suitable for the fast dynamics of impact wrench systems. Lastly, we demonstrate the effectiveness of the proposed framework through numerical simulations and hardware experiments. A statistical validation covering over 35,000 test trajectories confirms that the neural network maintains approximation error below 2\% throughout entire closed-loop trajectories in 97.99\% of cases, successfully replicating the optimal MPC policy. The results show that the NN controller consistently respects safety constraints in simulation, a critical task that the baseline speed controller repeatedly fails to achieve. Hardware experiments further validate the NN controller's performance and ability to operate within safe bounds on a real-world testbed. To the best of our knowledge, this work provides the first application of GP-augmented MPC with neural network approximation for impact wrench control on embedded hardware.

\subsubsection*{Related Work}
Control methods for impact wrench tightening procedures have been extensively investigated~\cite{bralla_iw_control,profunser_iw_control,ml_iw_control,posFb_iw_control,boltTight_iw_control}. In~\cite{profunser_iw_control}, target spindle speeds are set based on measurements from anvil and spindle speed sensors, with impacts identified by detecting speed differences between components. Similarly,~\cite{bralla_iw_control} employs impact and anvil position sensors to estimate angular progress and determine the continuation of operations. Regardless of the fastening procedure, motor torque commands typically rely on classical control strategies such as PID controllers for rotor speed tracking~\cite{bldc_mot_iw,bolt_tightening_mffc}. Other works leverage the linear relationship between clamping force and output torque~\cite{survey_automated_fastener}, employing learning-based strategies~\cite{bolt_tight_ctrl,fuzzyNN} and sliding-mode control~\cite{sliding_mode}. However, these approaches neither address vibration constraints in their formulation nor optimize performance during rotary impacts.

Model-based predictive control relies fundamentally on accurate system models to achieve high performance. Recent developments have addressed model uncertainty through data-driven augmentation, particularly via Gaussian process regression. Gaussian processes enable the learning of residual dynamics from experimental data while providing probabilistic uncertainty quantification~\cite{GPsforML}. Several works have integrated GP learning into MPC formulations, where the learned residual dynamics serve to improve prediction accuracy, while the uncertainty estimates are incorporated for robustness~\cite{8909368,8768048}. For a comprehensive review of GP-based MPC methods and their theoretical foundations, we refer the reader to~\cite{scampicchio2025gaussianprocessesdynamicslearning}.

While GP-augmented MPC improves model accuracy, the computational burden of solving the resulting nonlinear optimization problem remains substantial, particularly for resource-constrained embedded systems. Explicit MPC precomputes the optimization problem offline via multi-parametric programming, yielding an exact optimal control law as a piecewise affine function over a partitioned state space for linear systems~\cite{BEMPORAD_explicitMPC}. For nonlinear systems, approximate solutions based on piecewise affine representations can be obtained~\cite{JOHANSEN_approxExplMPC}. However, the number of polyhedral regions grows exponentially with the problem dimension, creating scalability challenges for memory and search complexity.
Machine learning techniques offer alternative approximation methods. Neural networks have been explored for MPC approximation since early work such as \cite{PARISINI_NNMPC}, with recent schemes designed to preserve safety guarantees probabilistically~\cite{Hertneck2018} or deterministically~\cite{hose2023_apporximateWSafety}. While neural networks could directly learn residual dynamics end-to-end, we employ Gaussian process regression for model augmentation due to the data efficiency smooth predictions with well-calibrated extrapolation behavior beyond the training distribution. The GP-learned residuals are then incorporated into the MPC optimization, while a separate neural network approximates the resulting MPC policy to achieve real-time performance.

\subsubsection*{Outline} 
The remainder of this paper is organized as follows. In Section~\ref{sec:preliminaries}, we present the impact wrench mechanism, control objectives, and system model. In Section~\ref{sec:method}, we detail the control architecture and methodology. In Section~\ref{sec:num_results}, we report simulation and experimental results. Finally, in Section~\ref{sec:conclus}, we conclude and discuss future extensions.

%% file: Sections/System_overview.tex
\section{Preliminaries}
\label{sec:preliminaries}
In this section, we describe the ball-and-cam mechanism of the impact wrench and its operational principle. Subsequently, we introduce the nonlinear model formulation governing the hammer and spindle dynamics. Lastly, we formalize the operational requirements and safety constraints that motivate the adoption of MPC for this application.

\subsection{Mechanism of the impact wrench}
Impact wrenches employ a ball-and-cam mechanism to convert continuous motor rotation into periodic impacts. A DC motor, powered by a battery, drives a spindle shaft through a gearbox transmission. The spindle features two v-shaped grooves that accommodate steel balls, which transmit torque to the hammer, a component with mirrored v-shaped grooves that enable both rotational and axial motion. A preloaded spring positioned between the spindle and hammer maintains continuous contact with the balls during operation. 

During steady operation, the spindle continuously rotates the hammer until the hammer's drive lugs engage with the anvil lugs, momentarily arresting hammer rotation. Continued spindle rotation induces angular displacement between the spindle and hammer, forcing the balls to slide backward along the v-grooves. This motion lifts the hammer axially and compresses the spring. Subsequently, the spindle re-accelerates the hammer while the spring decompresses, propelling the hammer downward until it re-engages the anvil. This cycle repeats at frequencies of up to 40 impacts per second, corresponding to one impact every 27~ms, motivating a control sampling time of 1~ms.

\subsection{System Model}

In this work, we model the impact wrench dynamics between impacts with a rigid-body model. The state of the model is given by $\boldsymbol{x} = [\phi_h, \phi_s, \omega_h, \omega_s]^T \in \mathbb{R}^4$, where $\phi_h$ and $\phi_s$ denote the hammer and spindle angular positions, respectively, and $\omega_h$ and $\omega_s$ denote the corresponding angular velocities. The system is controlled via the motor torque, summarized as input $u = \tau_m \in \mathbb{R}$. The states evolve according to the nonlinear continuous-time dynamics
\begin{align}
	\label{eq:ct-system}
    \dot{\boldsymbol{x}}(t) = f(\boldsymbol{x}(t), u(t), \boldsymbol{\theta}) + g(\boldsymbol{x}(t), u(t)),
\end{align}
where $f$ denotes the nominal model, $g$ the residual model accounting for systematic unmodeled effects, which we will learn from the data.

The nominal model $f$ comprises kinematic relationships for the angular positions ($\dot{\phi}_h = \omega_h$, $\dot{\phi}_s = \omega_s$) and dynamic equations governing the angular accelerations. The acceleration dynamics are linear in the parameter vector $\boldsymbol{\theta} = [\lambda, k_f x_0, k_f, 1]^T \in \mathbb{R}^4$, where $\lambda$ represents the input torque gain, $k_f x_0$ the spring preload, $k_f$ the spring stiffness, and the final element captures a bias term. This linear parameterization enables the structure
\begin{align}
	\label{eq:linear-in-params}
    \begin{bmatrix} \dot{\omega}_h \\ \dot{\omega}_s \end{bmatrix} = \boldsymbol{\Phi}(\boldsymbol{x}, u)^T \boldsymbol{\theta},
\end{align}
where $\boldsymbol{\Phi}(\boldsymbol{x}, u) \in \mathbb{R}^{4 \times 2}$ denotes the regressor matrix derived from the ball-and-cam mechanism kinematics and spring dynamics. For further details on impact wrench modeling, the reader is referred to~\cite{SysLevelMod}.

The residual model $g$ is constructed via Gaussian process (GP) regression to capture modeling errors and unmodeled dynamics not captured by the nominal model. The GP training procedure and feature selection are detailed in Section~\ref{subsubsec:res_dyn}.

The control formulation excludes two aspects of the system dynamics. First, upon impact, the hammer undergoes an impulsive state reset with post-impact velocity determined by the coefficient of restitution (COR), bounded in $[0,1]$. Since the COR is unknown a priori and varies during operation, reliable prediction of impact dynamics is infeasible and these effects are excluded from the model. Second, the anvil dynamics are not modeled as they are decoupled from the controlled system states and cannot be influenced by the motor torque input.

\subsection{Operational requirements}
\label{subsec: oper_req}
\begin{figure*}[t]
    \centering
    \subfloat[Proper operation]{%
        \includegraphics[width=0.48\textwidth]{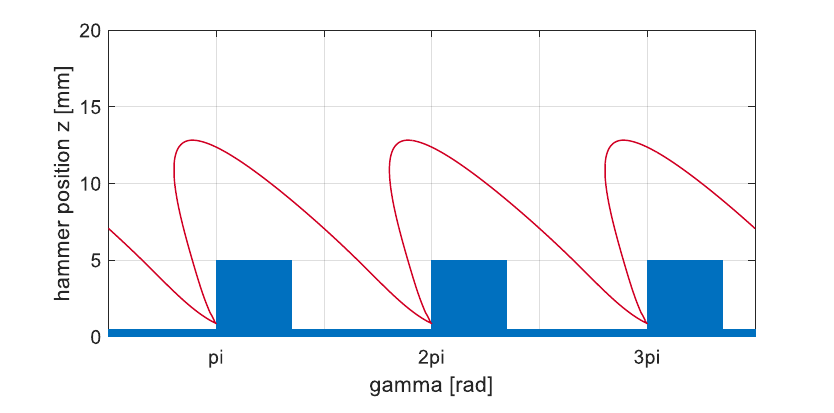}%
        \label{fig:proper_op}%
    }
    \hfill
    \subfloat[Improper operation]{%
        \includegraphics[width=0.48\textwidth]{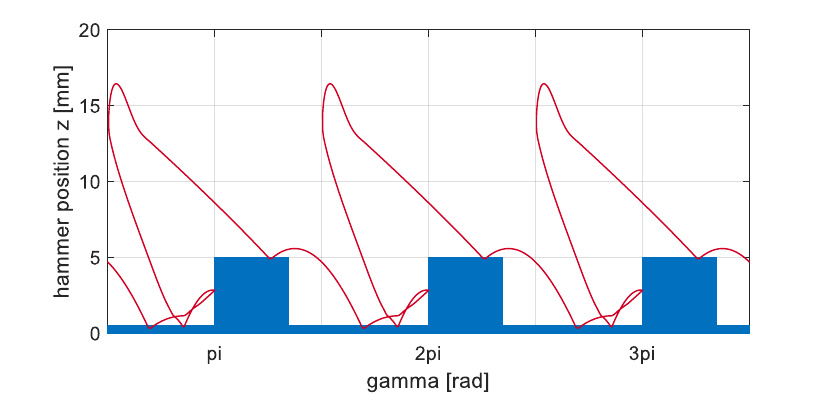}%
        \label{fig:improper_op}%
    }
    \caption{Impact wrench operational modes. Red line: hammer's lowest position; blue region: anvil. (a) Desirable impact configuration with direct hammer-to-anvil foot contact. (b) Failure mode exhibiting premature axial contact at multiple anvil surfaces, resulting in reduced torque output, increased component wear, and excessive vibrations.}
    \label{fig:impact_behavior}
\end{figure*}

Optimal impact wrench performance requires precise alignment between the hammer and anvil arms at impact to maximize contact area and minimize concentrated stresses that accelerate material fatigue, as illustrated in Fig.~\ref{fig:proper_op}. Conversely, axial contact between the components (Fig.~\ref{fig:improper_op}) dissipates kinetic energy, reducing output torque and degrading performance. Additionally, excessive hammer rebounds that reach the v-groove endpoints generate vibrations harmful to both operator safety and component longevity. These failure modes are exacerbated when non-standard sockets with elevated inertia are employed beyond the tool's design specifications. As depicted in Fig.~\ref{fig:impact_behavior}, successful operation requires maximizing hammer-anvil overlap at impact while preventing hammer rebound to the groove endpoints throughout the 27~ms impact cycle. The following subsection formalizes these operational requirements as mathematical constraints on the system dynamics.

\subsection{System Constraints}

The operational requirements from the previous subsection are formalized as mathematical constraints on the system dynamics~\eqref{eq:ct-system}. The hammer's axial position is determined by the spring angle $\phi_{\text{spring}}(t) = \phi_h(t) - \phi_s(t)$, representing the angular displacement between the hammer and spindle. To prevent v-groove endpoint collisions, the spring angle must remain within physical limits:
\begin{equation}
	\label{eq:spring_angle_lim}
	-\phi_b \leq \phi_h(t) - \phi_s(t) \leq \phi_b, \quad \forall t,
\end{equation}
where $\phi_b$ denotes the maximum allowable spring angle before endpoint contact.

Proper hammer-anvil alignment at impact time $t_{\text{imp}}$ requires the hammer to complete a $180°$ rotation plus the anvil's angular advancement from the previous impact. Let $\phi_{h,\text{pi}}$ denote the hammer's angular position at the previous impact, and $\phi_{ap}$ the anvil's angular advancement. The alignment constraint is:
\begin{equation}
	\label{eq:ham_at_imp}
	\phi_h(t_{\text{imp}}) = \phi_{h,\text{pi}} - \pi + \phi_{ap} \triangleq \phi_{h, \text{ref}}.
\end{equation}
Since measuring $\phi_{ap}$ requires an additional sensor, it is estimated within the control framework. Optimal impact energy occurs at spring angle $\phi_{\text{spring}}(t_{\text{imp}})$, yielding the spindle position constraint:
\begin{equation}
	\label{eq:spin_at_imp}
	\phi_s(t_{\text{imp}}) = \phi_{h, \text{ref}} - \phi_{\text{spring}}(t_{\text{imp}}).
\end{equation}

Finally, the motor torque is bounded by the physical actuator limits:
\begin{equation}
    \label{eq:torque_constr}
	u_{\text{min}} \leq u(t) \leq u_{\text{max}}, \quad \forall t.
\end{equation}

These constraints define the safe operating envelope and performance requirements that the controller must satisfy in real time.

%% file: Sections/Methodology.tex
\section{Methodology}
\label{sec:method}
In the following, we present the proposed control architecture, which comprises three key components: a free-final-time MPC formulation that handles variable impact timing, learning-based system identification and model augmentation via Gaussian process regression, and neural network approximation for real-time embedded deployment.

\subsection{Free Final-Time Model Predictive Control}

\subsubsection{Discretization}

The system is controlled exclusively between impact occurrences. Since impact timing varies with applied torque and plant state at the beginning of each impact period, influenced by the coefficient of restitution, the operation exhibits non-periodic behavior. A free-final-time predictive control strategy addresses this variability by fixing the prediction horizon while treating the discretization step-size $t_s \in \mathbb{R}_+$ as an optimization variable. As the system approaches impact, the step size decreases adaptively. This formulation follows the free-final-time class of optimization problems from optimal control theory~\cite{dpoc}.

Discretizing the system dynamics~\eqref{eq:ct-system} via the forward Euler method with step size $t_s$ yields:
\begin{align}
	\label{eq:dt-system}
	\boldsymbol{x}_{k+1} &= \bar{f}(\boldsymbol{x}_{k}, u_{k}, \boldsymbol{\theta}, t_s) + \bar{g}(\boldsymbol{z}_{k}, t_s) \\
    &= \boldsymbol{x}_{k} + t_s \left( f(\boldsymbol{x}_{k}, u_{k}, \boldsymbol{\theta}) + g(\boldsymbol{z}_{k}) \right), \notag
\end{align}
where $\bar{f}$ and $\bar{g}$ denote the discretized nominal and residual dynamics, respectively, and $\boldsymbol{z}$ represents the feature vector for the GP. Although higher-order integration methods such as RK4 offer improved accuracy, the forward Euler method is selected for its computational simplicity.

\subsubsection{Free Final-Time MPC Formulation}
At each sampling time $T_s = 1$ ms, the following optimization problem is solved:
\begin{subequations}
\label{subsec: mpc}
\begin{align}
    \min_{\boldsymbol{x}_{\cdot|k},\boldsymbol{u}_{\cdot|k},t_{s}} \quad & 
        J_u + J_s \label{mpc:cost} \\
    \textrm{s.t.} \quad 
      & \boldsymbol{x}_{0|k} = \boldsymbol{\hat{x}}(k), \label{mpc:init_cond}\\
      & \boldsymbol{x}_{1|k} = \bar{f}(\boldsymbol{x}_{0|k},u_{0|k},\boldsymbol{\hat{\theta}},T_{s}) + \bar{g}(\boldsymbol{z}_{0|k}, T_s), \label{mpc:dyn_eq_1st_step}\\
      & \boldsymbol{x}_{i+1|k} = \bar{f}(\boldsymbol{x}_{i|k},u_{i|k},\boldsymbol{\hat{\theta}}, t_{s}) + \bar{g}(\boldsymbol{z}_{i|k}, t_s), \label{mpc:dyn_eq}\\
      & \mathbf{A}\boldsymbol{x}_{i|k} \leq \mathbf{b} + \epsilon_1,  \label{mpc:state_constr}\\
       &  \mathbf{A}_f\boldsymbol{x}_{N|k} = \mathbf{b}_f + \epsilon_2, \label{mpc:terminal_constr}\\
      & \mathbf{C}u_{i|k} \leq \mathbf{d}, \label{mpc:input_constr}\\
      & 0 \leq t_{s} \leq t_{s,\text{max}}, \label{mpc:step_bounds}\\
      & \epsilon_{1,2} \geq 0, \label{mpc:lb_slack}\\
      & i \in \mathbb{N}_{0:N-1}, \notag
\end{align}
\end{subequations}
where subscript $_{i|k}$ denotes the $i$-th prediction from the initial condition at discrete time $k$, and $N$ is the prediction horizon. The cost function~\eqref{mpc:cost} comprises an input cost term $J_u$ and a slack penalty term $J_s$. The input cost $J_u$ penalizes deviations from the previous optimal input, $u^*_{0|k-1}$, and input rate changes, while the slack penalty $J_s$ softens state and terminal constraints:
\begin{align}
    J_u &= \lambda_u (u_{0|k}-u^*_{0|k-1})^2 + \sum_{i=1}^{N-1}\lambda_u (u_{i|k}-u_{i-1|k})^2,  \\
    J_s &= \lambda_{\epsilon_1}\epsilon_1^1 + \lambda_{\epsilon_2}\epsilon_2^2. \notag \label{eq:cost_slack}
\end{align}

The optimization problem is initialized at each step via \eqref{mpc:init_cond} using the current state estimate $\boldsymbol{\hat{x}}(k)$ from the EKF.

The system evolution is propagated by the dynamics \eqref{mpc:dyn_eq_1st_step} and \eqref{mpc:dyn_eq}, which use the discretized nominal model $\bar{f}$ augmented with a learned residual term $\bar{g}$ to compensate for model mismatch. A critical distinction exists: \eqref{mpc:dyn_eq_1st_step} employs the fixed sampling time $T_s = 1$ ms to ensure the optimal input $u_{0|k}^*$ is directly applied to the physical system without temporal averaging, while \eqref{mpc:dyn_eq} uses the free optimization variable $t_s$ for prediction steps $i \geq 1$. This enables adaptive adjustment of the internal prediction time step to satisfy terminal conditions.

Physical limits are enforced on states and inputs. To prevent end-of-groove hits, \eqref{mpc:state_constr} imposes the spring angle limits from~\eqref{eq:spring_angle_lim}, with matrices $\mathbf{A}$ and $\mathbf{b}$ encoding this condition. Similarly, \eqref{mpc:input_constr} enforces the motor torque limits from~\eqref{eq:torque_constr}, defined by matrices $\mathbf{C}$ and $\mathbf{d}$.

The terminal condition \eqref{mpc:terminal_constr} enforces maximum hammer-anvil overlap, as specified by~\eqref{eq:ham_at_imp} and~\eqref{eq:spin_at_imp}; matrices $\mathbf{A}_f$ and $\mathbf{b}_f$ are formulated to enforce this. Both the state and terminal conditions are softened via slack variables $\epsilon_1, \epsilon_2$ to prevent infeasibility.

Finally, \eqref{mpc:step_bounds} bounds the free step size $t_s$ to ensure forward time integration and numerical stability, while \eqref{mpc:lb_slack} enforces the non-negativity of the slack variables.

The optimization yields the optimal control policy $\pi_{\text{MPC}}$, computed at 1 kHz. Following each impact, the terminal reference~\eqref{eq:ham_at_imp} is updated with the new hammer position to define the target for the subsequent impact. However, if the predicted time to impact $Nt_s^*$ is very short (less than 2 ms) and the reference has not yet been updated due to delayed impact detection, the MPC optimization may converge to an incorrect solution based on the outdated reference. To ensure reliable control during such transient periods, a moving average of previous control signals is applied for $Nt_s^* \leq 2$ ms until the impact is detected and the reference is properly updated.

\subsection{System Identification}
\subsubsection{Parameter Identification}
\label{subsubsec:param_id}
The nominal model~\eqref{eq:ct-system} depends on the parameter vector $\boldsymbol{\theta} = [\lambda, k_f x_0, k_f, 1]^T \in \mathbb{R}^4$, whose nominal values exhibit manufacturing tolerances. Accurate parameter knowledge is critical for the MPC controller, as parameter errors degrade control performance and may induce aggressive torque switching.

Parameter identification is formulated as an offline regression problem using experimental data collected during normal operation. Exploiting the linear-in-parameters structure~\eqref{eq:linear-in-params}, we focus on the spindle acceleration dynamics. For each time sample $k$ between impacts, the measurement equation is
\begin{equation}
\label{eq:param_regression}
    \dot{\omega}_{s,k}^{\text{meas}} = \boldsymbol{\phi}_s(\boldsymbol{x}_k, u_k)^T \boldsymbol{\theta} + \nu_k, \notag
\end{equation}
where $\dot{\omega}_{s,k}^{\text{meas}} $ denotes the measured spindle angular acceleration, $\boldsymbol{\phi}_s(\boldsymbol{x}_k, u_k) \in \mathbb{R}^4$ represents the spindle row of the regressor matrix $\boldsymbol{\Phi}$ from~\eqref{eq:linear-in-params}, and $\nu_k$ captures measurement noise and modeling errors. Only spindle dynamics are employed, as spindle velocity is directly measured while hammer velocity must be estimated, enabling a more reliable computation of acceleration.

Data from multiple impact cycles are collected to construct the regression problem. Linear regression is employed to estimate the parameters by solving the least-squares problem.
\begin{equation}
    \boldsymbol{\hat{\theta}} = \arg\min_{\boldsymbol{\theta}} \sum_{k=1}^{K} \left(\dot{\omega}_{s,k}^{\text{meas}}  - \boldsymbol{\phi}_s(\boldsymbol{x}_k, u_k)^T \boldsymbol{\theta}\right)^2, \notag
\end{equation}
where $K$ denotes the total number of samples collected. The identified parameters $\boldsymbol{\hat{\theta}} = [\hat{\lambda}, \widehat{k_f x_0}, \hat{k}_f, 1]^T$ are subsequently used in the MPC formulation~\eqref{subsec: mpc}. Despite this identification procedure, residual model mismatch persists due to unmodeled dynamics, motivating the GP augmentation described next.

\subsubsection{Residual Dynamics Modeling}
\label{subsubsec:res_dyn}
The goal of learning the residual dynamics $g$ is to account for effects not captured by the nominal model $f$, thereby improving predictive performance. For controllers that rely on model propagation, such as MPC, predictive accuracy strongly impacts control performance~\cite{9981780}.

We employ Gaussian process (GP) regression~\cite{Rasmussen2006Gaussian} to learn residual prediction errors from experimental data. The learned residual model focuses on spindle angular acceleration. Consequently, the residual model has the structure
\begin{equation}
\label{eq:gp_residual}
    g(\boldsymbol{z}_k) = \begin{bmatrix} 0 \\ 0 \\ 0 \\ \mu_{\text{GP}}(\boldsymbol{z}_k) \notag \end{bmatrix},
\end{equation}
where $\boldsymbol{z}_k = [\phi_{spring, k},\dot{\phi}_{spring, k}, u_k]^T \in \mathbb{R}^3$ denotes the feature vector consisting of the spring angle, spring velocity, and motor torque, and $\mu_{\text{GP}}(\boldsymbol{z}_k): \mathbb{R}^3 \rightarrow \mathbb{R}$ is the GP mean function, which maps the feature vector to the predicted spindle acceleration residual.

Training data are collected offline from experimental trials, where we compute spindle acceleration directly from raw sensor measurements. Specifically, spindle velocity measurements are numerically differentiated to obtain acceleration, employing non-causal filtering techniques to reduce measurement noise and obtain reliable ground truth.
The training targets are computed as
\begin{equation}
\label{eq:gp_training_target}
    e_{\omega_s,k} = \dot{\omega}_{s,k}^{\text{meas}} - \boldsymbol{\phi}_s(\boldsymbol{x}_k, u_k)^T \boldsymbol{\hat{\theta}}, \notag
\end{equation}
where $\dot{\omega}_{s,k}^{\text{meas}}$ denotes the measured acceleration obtained from filtered numerical differentiation, and $\boldsymbol{\phi}_s$ represents the spindle row of the regressor matrix from~\eqref{eq:linear-in-params}. Data are collected from multiple impact cycles across varying operating conditions to ensure representative coverage of the state-input space. Given the large data volume, we apply an active learning strategy~\cite{DBLP:journals/corr/abs-1901-05954} that combines uncertainty sampling with spatial diversity via clustering, yielding a compact training set while maintaining feature space coverage.

The learned mean function $\mu_{\text{GP}}(\boldsymbol{z}_k)$ provides point predictions during online operation, which are integrated into the MPC dynamics constraints~\eqref{mpc:dyn_eq_1st_step}--\eqref{mpc:dyn_eq}.

\subsection{Learning-Based MPC Approximation}
\label{subsec:nn_approximation}

The MPC optimization problem~\eqref{subsec: mpc} must execute at 1~kHz to achieve the desired control performance. However, solving the nonlinear program exceeds the 1 ms computational budget available on the embedded microcontroller. To address this bottleneck, we approximate the MPC policy using a feedforward neural network trained via supervised learning on optimal MPC solutions.

\subsubsection{Data Generation and Network Training}

Training data are generated offline by solving the MPC problem~\eqref{subsec: mpc} across the feasible operating region. Initial states, impact references, and previous control inputs are sampled uniformly from their respective feasible ranges. For each sample, the augmented input $\boldsymbol{\xi} = [\boldsymbol{x}_0^T, \phi_{h,\text{ref}}, u_{\text{prev}}]^T \in \mathbb{R}^6$ is constructed, and the corresponding optimal solution $[u^*, t_s^*]^T$ is computed.

\subsubsection{Closed-Loop Validation with Statistical Guarantees}

Rather than relying on single-step prediction accuracy, we evaluate the approximation quality through closed-loop simulation with statistical confidence bounds, following the approach of~\cite{Hertneck2018}. This validation captures error accumulation over time and provides probabilistic guarantees on controller performance.

For validation, we consider trajectories $X_i$ of the closed-loop system starting from initial conditions $\boldsymbol{x}_i(0)$ sampled independently from the feasible region. Each trajectory evolves under neural network control, with the state updated at 1~ms intervals using the learned dynamics~\eqref{eq:dt-system}. At each control step along the trajectory, we compute both the NN prediction $\boldsymbol{\pi}_{\text{NN}}(\boldsymbol{x}_k)$ and the optimal MPC solution $\boldsymbol{\pi}_{\text{MPC}}(\boldsymbol{x}_k)$ from the same state, measuring the normalized approximation error.
We define an indicator function
\begin{equation}
\label{eq:indicator}
    I(X_i) = \begin{cases}
        1 & \text{if } \max_{k} \left\| \frac{\boldsymbol{\pi}_{\text{NN}}(\boldsymbol{x}_k) - \boldsymbol{\pi}_{\text{MPC}}(\boldsymbol{x}_k)}{\boldsymbol{y}_{\max} - \boldsymbol{y}_{\min}} \right\|_{\infty} < \eta, \, \forall \boldsymbol{x}_k \in X_i \\
        0 & \text{otherwise}
    \end{cases},
    \notag
\end{equation}
where $\eta$ is the error threshold and $\boldsymbol{y}_{\max} - \boldsymbol{y}_{\min}$ represents the output normalization range. A trajectory succeeds if the maximum element-wise normalized error remains below $\eta$ throughout the entire closed-loop trajectory.
Given $p$ independent test trajectories with initial conditions sampled from distribution $\Omega$ over the feasible region, the empirical success rate is
\begin{equation}
\label{eq:empirical_risk}
    \tilde{\mu} = \frac{1}{p} \sum_{i=1}^{p} I(X_i). \notag
\end{equation}
The true success probability $\mu = P[I(X_i) = 1]$ can be bounded using Hoeffding's inequality:
\begin{equation}
\label{eq:hoeffding}
    P[|\tilde{\mu} - \mu| \geq \epsilon_h] \leq 2\exp(-2p\epsilon_h^2), \notag
\end{equation}
where $\epsilon_h = \sqrt{-\ln(\delta_h/2)/(2p)}$ and $\delta_h$ are the confidence parameters. This implies that with confidence at least $1 - \delta_h$, the true success probability satisfies $\mu \geq \tilde{\mu} - \epsilon_h$.

The validation succeeds if the statistical lower bound exceeds a critical threshold $\mu_{\text{crit}}$:
\begin{equation}
\label{eq:validation_criterion}
    \mu_{\text{crit}} \leq \tilde{\mu} - \epsilon_h.\notag
\end{equation}

%% file: Sections/Numerical_Results.tex
\section{Numerical Results}
\label{sec:num_results}
We evaluate the proposed control architecture through simulation and hardware experiments. Simulation results compare a baseline speed controller, exact MPC, and the NN-approximated MPC in terms of computational efficiency and control performance. Hardware experiments validate the complete pipeline on the physical impact wrench, demonstrating real-world constraint satisfaction.

\subsection{Simulation Experiments}
\subsubsection{Experimental Setup}
Simulations use the continuous-time model~\eqref{eq:ct-system} with parameters $\boldsymbol{\hat{\theta}}$ identified via the procedure in Section~\ref{subsubsec:param_id}. Table~\ref{tab:mpc_params} summarizes the MPC configuration, and Table~\ref{tab:nn_params} details the neural network hyperparameters. 

As established in our operational requirements (see Section~\ref{subsec: oper_req} and Fig.~\ref{fig:impact_behavior}), the spring bound $\phi_b = 2.11$ rad is a critical constraint. It is directly correlated to the maximum hammer position $x_{\text{ham}} = 0.012$ m, which is the physical end of the V-groove. Exceeding this limit causes the undesired rebounds and performance degradation (Fig.~\ref{fig:improper_op}) that the controller is designed to prevent.

The MPC problem~\eqref{subsec: mpc} is solved using CasADi~\cite{Andersson2019} with the IPOPT solver. All timing measurements are performed on an Intel(R) Core(TM) i9-7940X CPU @ 3.10 GHz.

\begin{table}[t]
\centering
\caption{MPC configuration parameters.}
\label{tab:mpc_params}
\begin{tabular}{lc}
\hline
Parameter & Value \\
\hline
Prediction horizon $N$ & 30 \\
Sampling time $T_s$ [ms] & 1.0 \\
Input weight $\lambda_u$ & 4.0 \\
Slack weights $\lambda_{\epsilon_1}, \lambda_{\epsilon_2}$ & [100, 5] \\
Input bounds $u_{\min}, u_{\max}$ [Nm] & [0, -0.5] \\
Spring bound $\phi_b$ [rad] & 2.11 \\
$\phi_{\text{spring}}(t_{\text{imp}})$ & 0.2 \\
Max step size $t_{s,\text{max}}$ [ms] & 1.0 \\
\hline
\end{tabular}
\end{table}

\begin{table}[t]
\centering
\caption{Neural network hyperparameters.}
\label{tab:nn_params}
\begin{tabular}{lc}
\hline
Parameter & Value \\
\hline
Architecture & 5 × 50 (Tanh) \\
Training samples & 1.5M \\
Optimizer & AdamW \\
Learning rate & 5e-3 (plateau) \\
Batch size & 256 \\
\hline
\end{tabular}
\end{table}

\subsubsection{Neural Network Approximation Accuracy}
Table~\ref{tab:statistical_validation} presents the closed-loop statistical validation results from Section~\ref{subsec:nn_approximation}. The NN achieves a 97.99\% success rate across 35{,}000 test trajectories, demonstrating closed-loop performance very close to the MPC baseline. With a statistical lower bound of $\tilde{\mu} - \epsilon_h = 0.971$ exceeding the critical threshold $\mu_{\text{crit}} = 0.970$ at 99\% confidence, the validation criterion is satisfied. Successful trajectories maintain sub-0.1\% mean errors throughout execution, validating the NN's ability to replicate MPC behavior in the vast majority of operating conditions.

\begin{table}[t]
\centering
\caption{Statistical validation results over 35{,}000 test trajectories.}
\label{tab:statistical_validation}
\begin{tabular}{lc}
\hline
Metric & Value \\
\hline
Test trajectories $p$ & 35{,}000 \\
Empirical success rate $\tilde{\mu}$ & 0.9799 \\
Confidence level $1 - \delta_h$ & 0.99 \\
Statistical margin $\epsilon_h$ & 0.0087 \\
Lower bound $\tilde{\mu} - \epsilon_h$ & 0.9712 \\
Critical threshold $\mu_{\text{crit}}$ & 0.9700 \\
Error threshold $\eta$ & 2\% \\
Mean error (motor torque) & 0.093\% $\pm$ 0.086\% \\
Mean error (time variable) & 0.025\% $\pm$ 0.018\% \\
\hline
\end{tabular}
\end{table}

\subsubsection{Controller Performance Comparison} 
\begin{figure*}[t]
    \centering
    \subfloat[Spring Angle at Impact]{%
        \includegraphics[width=0.48\textwidth]{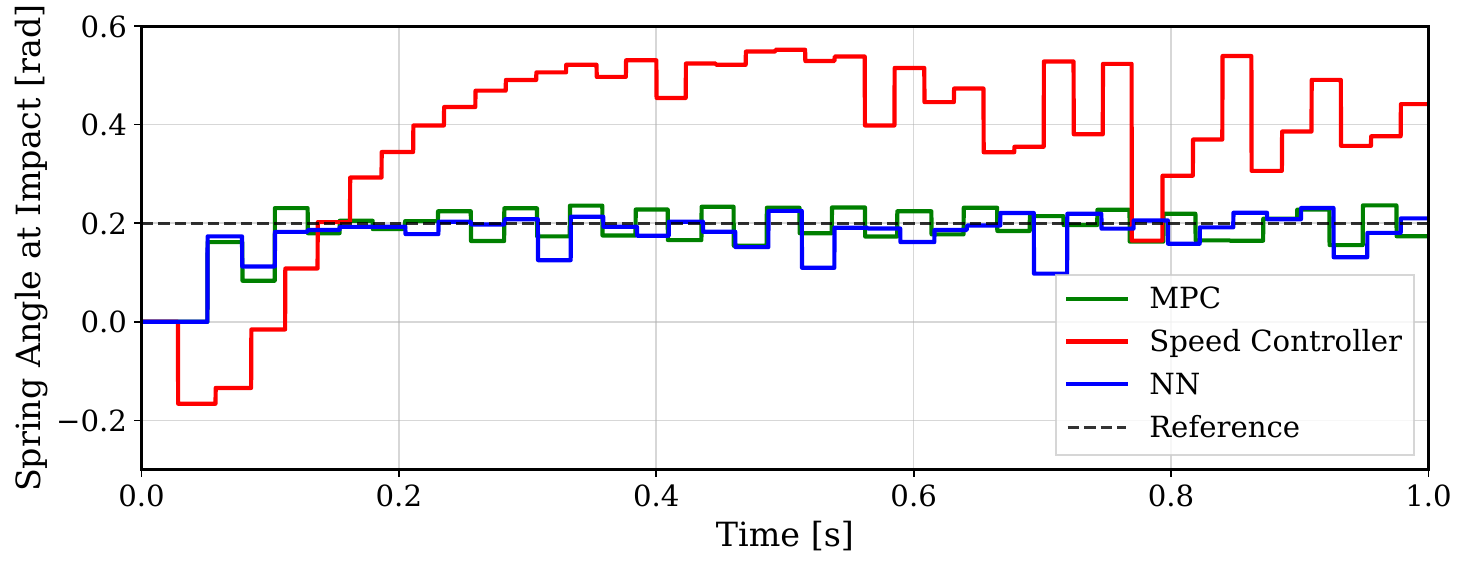}%
        \label{fig:spring_angle_comp}%
    }
    \hfill
    \subfloat[Max Hammer X Position]{%
        \includegraphics[width=0.48\textwidth]{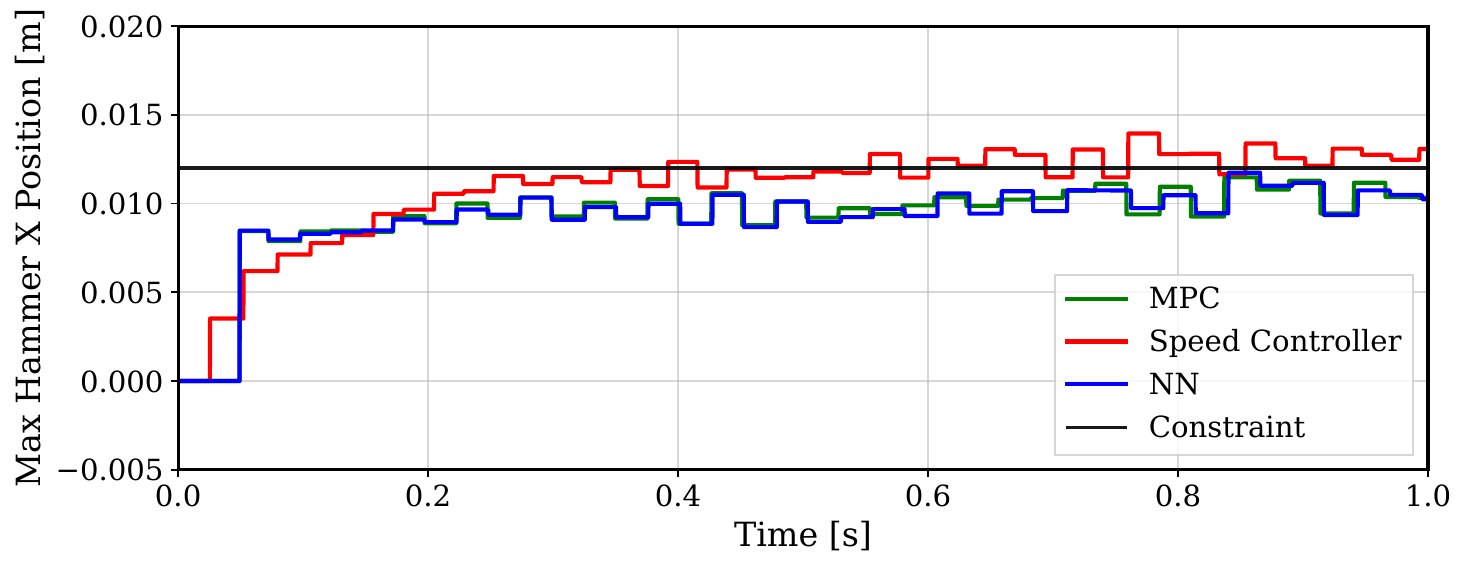}%
        \label{fig:hammer_pos_comp}%
    }
    \caption{Simulation comparison of the Neural Network (NN), Model Predictive Control (MPC), and a baseline speed controller. (a) Spring Angle at Impact: The NN (blue line) and MPC (green line) controllers track the 0.2 rad reference (black dashed). (b) Max Hammer X Position: Both the NN and MPC controllers respect the 0.012 m constraint (black solid), while the speed controller violates it.}
    \label{fig:closed_loop_comparison}
\end{figure*}

Figure \ref{fig:closed_loop_comparison} compares the closed-loop trajectories of the baseline speed controller, the exact MPC, and the NN-approximated MPC over 40 consecutive impact cycles.

Both the exact MPC and the NN controller successfully maintain constraint satisfaction. In contrast, the speed controller (red line) exhibits significant and repeated constraint violations, demonstrating the clear superiority of the predictive control strategy.

Specifically, Figure~\ref{fig:spring_angle_comp} shows the NN and exact MPC closely tracking the 0.2 rad spring angle at impact reference. Figure~\ref{fig:hammer_pos_comp} confirms they both respect the 0.012 m maximum hammer position constraint. The speed controller fails in both metrics.

\subsubsection{Computational Efficiency}
Table~\ref{tab:computation} demonstrates that the NN approximation enables real-time execution, while exact MPC consistently violates the 1 ms control cycle requirement.

\begin{table}[t]
\centering
\caption{Computational performance over 35{,}000 trajectories.}
\label{tab:computation}
\begin{tabular}{lccc}
Method & Mean [ms] & Std [ms] & Max [ms] \\
\midrule
MPC (IPOPT) & 115.9 & 80.0 & 326 \\
NN Approximation & 0.24 & 0.03 & 1.64 \\
Speedup & 490$\times$ & -- & -- \\

\end{tabular}
\end{table}

\subsection{Hardware Experiments}
\subsubsection{Experimental Setup}
The control algorithm was validated on an industrial impact wrench testbed. 

The experimental platform consists of an embedded microcontroller
that executes the complete control pipeline (EKF state estimation, 
GP prediction, NN-based MPC) at a 1~kHz 
control frequency. The controller sends torque commands to a 
commercial-off-the-shelf impact wrench. The wrench is 
instrumented with a motor angle sensor and a 
hammer angle sensor, which provide high-speed feedback to the 
state estimator. 

\subsubsection{Hardware Validation and Controller Transition} 
\begin{figure*}[t]
    \centering
    \subfloat[Spring Angle at Impact]{%
        \includegraphics[width=0.48\textwidth]{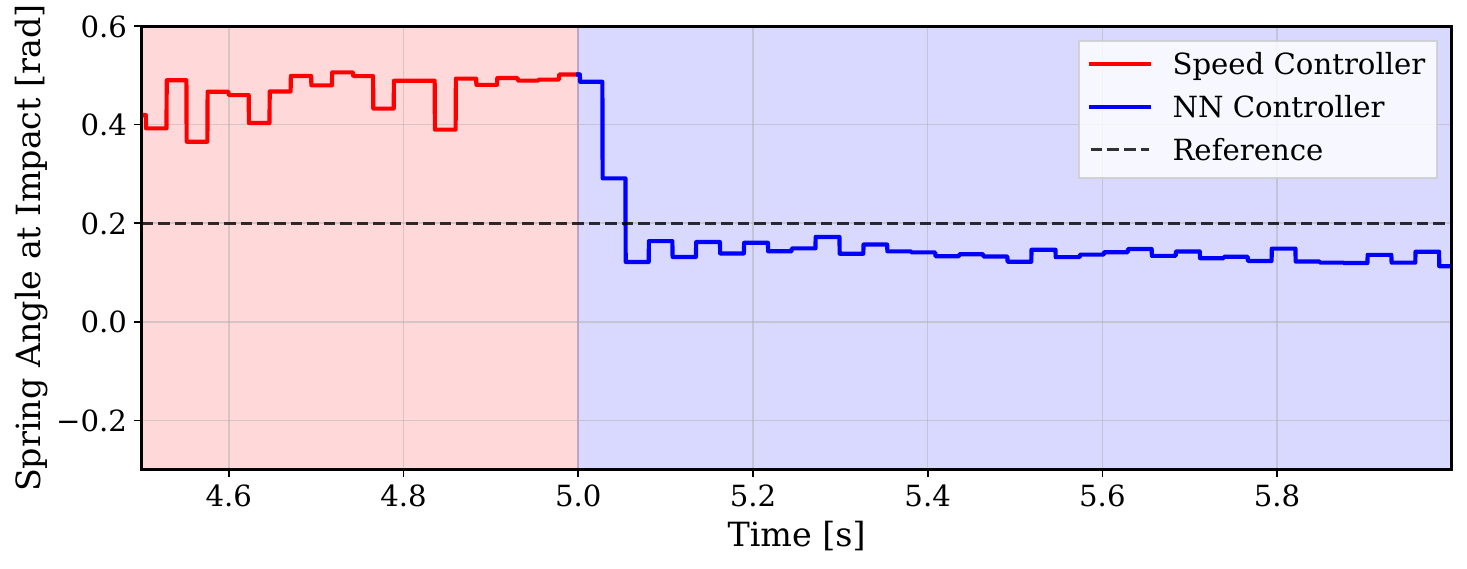}%
        \label{fig:hardware_spring}%
    }
    \hfill
    \subfloat[Max Hammer X Position]{%
        \includegraphics[width=0.48\textwidth]{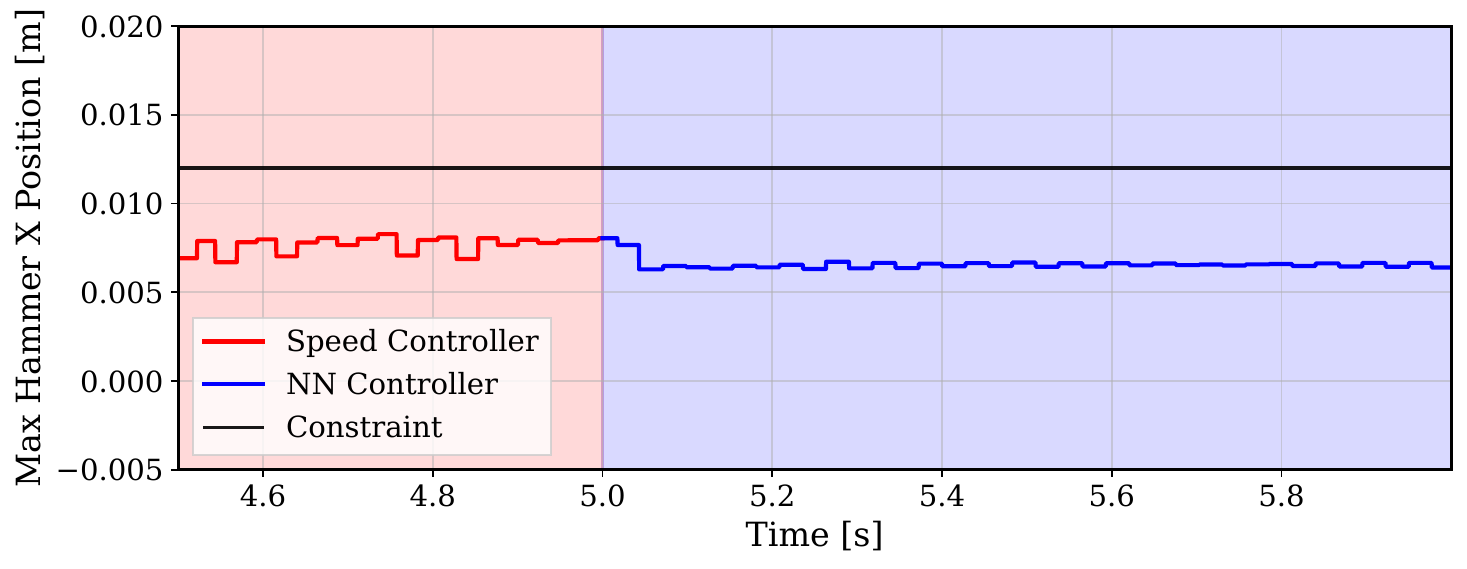}%
        \label{fig:hardware_hammer}%
    }
    \caption{Hardware experimental results showing constraint satisfaction during controller transition. Red shaded region: speed Controller (4.5--5.0s); Blue shaded region: NN Controller (5.0--6.0s). (a) Spring angle tracking demonstrates improved reference following under NN. (b) Maximum hammer position remains within safety constraints throughout both control phases.}
    \label{fig:hardware_results}
\end{figure*}

The NN-based controller's performance is validated on the hardware testbed, as shown in Figure~\ref{fig:hardware_results}. The experiment demonstrates a live transition from the baseline speed controller (active from 4.5~s to 5.0~s, red region) to the proposed NN controller (active from 5.0~s onward, blue region).

Figure~\ref{fig:hardware_spring} shows the spring angle at impact. The speed controller (red) shows highly variable performance. The moment the NN controller (blue) is activated at 5.0~s, the spring angle at impact is significantly better tracked. However, it converges to approximately 0.16~rad, exhibiting a small steady-state offset from the 0.2~rad reference. This offset likely indicates a slight model mismatch between the simulation environment used for training and the physical hardware, suggesting that unmodeled effects persist despite the GP-based model augmentation employed offline during controller development. An offset-free GP-augmented MPC scheme~\cite{carron2019data} can eliminate this residual offset.
Figure~\ref{fig:hardware_hammer} shows the maximum hammer position. In this specific experiment, the operational resistance was moderate, and as a result, neither the speed controller nor the NN controller violated the 0.012~m safety constraint. 

%% file: Sections/Conclusion.tex
\section{Conclusion}
\label{sec:conclus}
This paper presents a learning-augmented MPC architecture for impact wrench torque control that combines free-final-time optimization, Gaussian process residual learning, and neural network policy approximation. The NN-based controller achieves a 490$\times$ speedup over exact MPC, with sub-millisecond execution times that enable real-time 1 kHz control on embedded microcontrollers while maintaining a 97.99\% closed-loop success rate across diverse operating conditions. Hardware validation on the impact wrench confirms constraint satisfaction and stable operation. Future work will investigate parameter-adaptive neural networks that exploit MPC sensitivity information to rapidly adjust control policies across manufacturing variations without requiring network retraining.

\section*{Acknowledgment}
We thank Petar Stamenkovic for his support in the early phases of the project.

%% file: bibliography.bib
@article{SysLevelMod,
author = {Zhang, Shengli and Tang, J.},
title = {System-Level Modeling and Parametric Identification of Electric Impact Wrench},
journal = {Journal of Manufacturing Science and Engineering},
volume = {138},
number = {11},
pages = {111010},
year = {2016}
}

@article{peripheralnerveendings,
title = {Effects of power tool vibration on peripheral nerve endings},
author = {Jordan Zimmerman and James Bain and Magnus Persson and Danny Riley},
journal = {International Journal of Industrial Ergonomics},
volume = {62},
pages = {42-47},
year = {2017}
}

@article{Hand-armVibrationEffectsonPerformanceTactileAcuityandTemperatureofHand,
author = {Forouharmajd, Farhad and Yadegari, Mehrdad and Ahmadvand, Masoumeh and Pourabdian, Siamak},
title = {Hand-arm Vibration Effects on Performance, Tactile Acuity, and Temperature of Hand},
journal = {Journal of Medical Signals and Sensors},
volume = {7},
pages = {252-260},
year = {2017}
}

@article{EvaluationofImpactWrenchVibrationEmissionsandTestMethods,
author = {McDowell, Thomas W. and Dong, R. G. and Xu, X. and Welcome, D. E. and Warren, C.},
title = {An Evaluation of Impact Wrench Vibration Emissions and Test Methods},
journal = {The Annals of Occupational Hygiene},
volume = {52},
number = {2},
pages = {125-138},
year = {2008}
}

@article{InvestigCharacVibr,
title = {An investigation on characteristics of the vibration transmitted to wrist and elbow in the operation of impact wrenches},
author = {X.S. Xu and D.E. Welcome and T.W. McDowell and C. Warren and R.G. Dong},
journal = {International Journal of Industrial Ergonomics},
volume = {39},
number = {1},
pages = {174-184},
year = {2009}
}

@article{BEMPORAD_explicitMPC,
title = {The explicit linear quadratic regulator for constrained systems},
author = {Alberto Bemporad and Manfred Morari and Vivek Dua and Efstratios N. Pistikopoulos},
journal = {Automatica},
volume = {38},
number = {1},
pages = {3-20},
year = {2002}
}

@article{JOHANSEN_approxExplMPC,
title = {Approximate explicit receding horizon control of constrained nonlinear systems},
author = {Tor A. Johansen},
journal = {Automatica},
volume = {40},
number = {2},
pages = {293-300},
year = {2004}
}

@article{Hertneck2018,
title={Learning an Approximate Model Predictive Controller With Guarantees},
author={Hertneck, Michael and Kohler, Johannes and Trimpe, Sebastian and Allgower, Frank},
journal={IEEE Control Systems Letters},
volume={2},
number={3},
year={2018}
}

@misc{hose2023_apporximateWSafety,
title={Approximate non-linear model predictive control with safety-augmented neural networks},
author={Henrik Hose and Johannes Köhler and Melanie N. Zeilinger and Sebastian Trimpe},
year={2023}
}

@book{GPsforML,
author = {Rasmussen, Carl Edward and Williams, Christopher K. I.},
title = {Gaussian Processes for Machine Learning},
publisher = {The MIT Press},
year = {2005}
}

@book{dpoc,
author = {Bertsekas, Dimitri P.},
title = {Dynamic Programming and Optimal Control, Vol. II},
publisher = {Athena Scientific},
year = {2007}
}

@article{PARISINI_NNMPC,
title = {A receding-horizon regulator for nonlinear systems and a neural approximation},
journal = {Automatica},
volume = {31},
number = {10},
pages = {1443-1451},
year = {1995}
}

@patent{bralla_iw_control,
title = {Control method for an impact wrench},
author = {Alberding, Matthäus and Wötzl, Christian and Bralla, Dario},
number = {U.S. Patent 11465263B2},
year = {2022}
}

@patent{profunser_iw_control,
title = {Impact wrench and control method for an impact wrench},
author = {Profunser, Dieter and Brugger, Peter and Kuntner, Jochen},
number = {U.S. Patent 20100263890A1},
year = {2010}
}

@patent{ml_iw_control,
title = {Power tool including a machine learning block for controlling a seating of a fastener},
author = {Abbott, Jonathan E. and Evankovich, Justin A.},
number = {U.S. Patent 20250028289A1},
year = {2025}
}

@patent{posFb_iw_control,
title = {Position feedback control method and power tool},
author = {Reese, Brian Todd and Mayer, Cody Lyle},
number = {U.S. Patent 20240335917A1},
year = {2024}
}

@patent{boltTight_iw_control,
title = {Bolt-tightening method using an impact wrench},
author = {Noda, Hirotoshi},
number = {U.S. Patent 005457866A},
year = {1995}
}

@inproceedings{bolt_tight_ctrl,
author={Fujinaka, T. and Nakano, H. and Omatu, S.},
title={Bolt tightening control using neural networks},
booktitle={2001 IEEE International Conference on Systems, Man and Cybernetics},
volume={3},
pages={1390-1395},
year={2001}
}

@article{bldc_mot_iw,
author = {He, Chengyuan and Wu, Thomas},
title = {Permanent Magnet Brushless DC Motor and Mechanical Structure Design for the Electric Impact Wrench System},
journal = {Energies},
volume = {11},
number = {6},
year = {2018}
}

@article{bolt_tightening_mffc,
author={Deters, Christian and Lam, Hak-Keung and Secco, Emanuele Lindo and Würdemann, Helge A. and Seneviratne, Lakmal D. and Althoefer, Kaspar},
title={Accurate Bolt Tightening Using Model-Free Fuzzy Control for Wind Turbine Hub Bearing Assembly},
journal={IEEE Transactions on Control Systems Technology},
volume={23},
number={1},
pages={1-12},
year={2015}
}

@article{survey_automated_fastener,
author={Jia, Zhenzhong and Bhatia, Ankit and Aronson, Reuben M. and Bourne, David and Mason, Matthew T.},
title={A Survey of Automated Threaded Fastening},
journal={IEEE Transactions on Automation Science and Engineering},
volume={16},
number={1},
pages={298-310},
year={2019}
}

@article{fuzzyNN,
author = {Juang, Chia-Feng and Chen, Yi-Wei},
title = {Automatic Hitting-Duration Estimation of a Rechargeable Impact Wrench Using a Fuzzy Neural Network to Reach Target Toques},
journal = {International Journal of Fuzzy Systems},
volume = {25},
year = {2022}
}

@article{sliding_mode,
author = {Zhimin Wu and Guigang Zhang and Wenjuan Du and Jian Wang and Fengyang Han and Dianwei Qian},
title = {Torque control of bolt tightening process through adaptive-gain second-order sliding mode},
journal = {Measurement and Control},
volume = {53},
number = {7-8},
pages = {1131-1143},
year = {2020}
}

@inproceedings{9981780,
author={Fröhlich, Lukas P. and Küttel, Christian and Arcari, Elena and Hewing, Lukas and Zeilinger, Melanie N. and Carron, Andrea},
title={Contextual Tuning of Model Predictive Control for Autonomous Racing},
booktitle={2022 IEEE/RSJ International Conference on Intelligent Robots and Systems (IROS)},
pages={10555-10562},
year={2022}
}

@book{Rasmussen2006Gaussian,
author = {Rasmussen, Carl Edward and Williams, Christopher K. I.},
title = {Gaussian Processes for Machine Learning},
publisher = {The MIT Press},
year = {2006}
}

@article{8768048,
author={Carron, Andrea and Arcari, Elena and Wermelinger, Martin and Hewing, Lukas and Hutter, Marco and Zeilinger, Melanie N.},
title={Data-Driven Model Predictive Control for Trajectory Tracking With a Robotic Arm},
journal={IEEE Robotics and Automation Letters},
volume={4},
number={4},
pages={3758-3765},
year={2019}
}

@article{8909368,
author={Hewing, Lukas and Kabzan, Juraj and Zeilinger, Melanie N.},
title={Cautious Model Predictive Control Using Gaussian Process Regression},
journal={IEEE Transactions on Control Systems Technology},
volume={28},
number={6},
pages={2736-2743},
year={2020}
}

@article{DBLP:journals/corr/abs-1901-05954,
author = {Fedor Zhdanov},
title = {Diverse mini-batch Active Learning},
journal = {CoRR},
volume = {abs/1901.05954},
year = {2019}
}

@article{Andersson2019,
author = {Andersson, Joel and Gillis, Joris and Horn, Greg and Rawlings, James and Diehl, Moritz},
title = {CasADi: a software framework for nonlinear optimization and optimal control},
journal = {Mathematical Programming Computation},
volume = {11},
year = {2019}
}

@misc{scampicchio2025gaussianprocessesdynamicslearning,
title={Gaussian processes for dynamics learning in model predictive control},
author={Anna Scampicchio and Elena Arcari and Amon Lahr and Melanie N. Zeilinger},
year={2025}
}

@article{carron2019data,
  title={Data-driven model predictive control for trajectory tracking with a robotic arm},
  author={Carron, Andrea and Arcari, Elena and Wermelinger, Martin and Hewing, Lukas and Hutter, Marco and Zeilinger, Melanie N},
  journal={IEEE Robotics and Automation Letters},
  volume={4},
  number={4},
  pages={3758--3765},
  year={2019},
  publisher={IEEE}
}
